\documentclass{mn2e}
\usepackage{epsf}
\usepackage{times}

\title[]
      {ASMOOTH: A simple and efficient algorithm for adaptive kernel smoothing
       of two-dimensional imaging data}
\author[Ebeling et al.]
      {\parbox{\textwidth}{H.\ Ebeling$^{1,2}$, D.A.\ White$^{1,3}$, F.V.N.\ Rangarajan$^1$}\\ \\
       $^1$ Institute of Astronomy, Madingley Road, Cambridge CB3\,0HA, UK\\
       $^2$ Institute for Astronomy, 2680 Woodlawn Drive, Honolulu, HI 96822, USA\\
       $^3$ SunGard Trading and Risk Systems, Enterprise House, Suite 7, Vision Park,
            Chivers Way, Histon, Cambridge CB4 9ZR, UK}

\date{Received ***; in original form ***}
\begin{document}

\maketitle

\begin{abstract} 

An efficient algorithm for adaptive kernel smoothing (AKS) of two-dimensional
imaging data has been developed and implemented using the Interactive Data
Language (IDL). The functional form of the kernel can be varied (top-hat,
Gaussian etc.) to allow different weighting of the event counts registered
within the smoothing region. For each individual pixel the algorithm increases
the smoothing scale until the signal-to-noise ratio (s.n.r.) within the kernel
reaches a preset value.  Thus, noise is suppressed very efficiently, while at
the same time real structure, i.e.\ signal that is locally significant at the
selected s.n.r.\ level, is preserved on all scales. In particular, extended
features in noise-dominated regions are visually enhanced. The {\sc asmooth}
algorithm differs from other AKS routines in that it allows a quantitative
assessment of the goodness of the local signal estimation by producing
adaptively smoothed images in which all pixel values share the same
signal-to-noise ratio {\em above the background}.

We apply {\sc asmooth} to both real observational data (an X-ray image of
clusters of galaxies obtained with the Chandra X-ray Observatory) and to a
simulated data set. We find the {\sc asmooth}ed images to be fair
representations of the input data in the sense that the residuals are consistent
with pure noise, i.e.\ they possess Poissonian variance and a near-Gaussian
distribution around a mean of zero, and are spatially uncorrelated.

\end{abstract}


\section{Introduction} 

Smoothing of two-dimensional event distributions is a procedure routinely used
in many fields of data analysis. In practice, smoothing means the convolution
\begin{eqnarray}
   I'(\vec{r}\,) \equiv I(\vec{r}\,) \otimes {\cal K}(\vec{r}\,) 
                  = \int_{{\sf I\hspace*{-0.25ex}R}^2} 
                     {\cal K}(\vec{r}-\vec{r}\,')\, I(\vec{r}\,') 
                              \, d\vec{r}\,'  \label{convol} \\ 
   \left(  \int_{{\sf I\hspace*{-0.25ex}R}^2} {\cal K}(\vec{r}\,)\,d\vec{r}\,' 
                       = 1 \right) \hspace*{2cm}
   \nonumber
\end{eqnarray}
of the measured data $I(\vec{r}\,)$ with a kernel function $\cal K$ (often also
called `filter' or `window function'). Although the raw data may be an image in
the term's common meaning [i.e. the data set can be represented as a function
  $I(x,y)$ where $I$ is some intensity, and $x$ and $y$ are spatial
  coordinates], the two coordinates $x$ and $y$ can, in principle, describe any
two-dimensional parameter space.  The coordinates $x$ and $y$ are assumed to
take only discrete values, i.e.\ the events are binned into $(x,y)$
intervals. The only requirement on $I$ that we shall assume in all of the
following is that $I$ is the result of a counting process in some detector, such
that $I(x,y) \in {\sf I\hspace*{-0.25ex}N}_0$.

An image, as defined above, is a two-dimensional histogram and is thus often a
coarse representation of the underlying probability density distribution
(e.g.\ Merrit \& Tremblay 1994, Vio et al.\ 1994). However, for certain
experiments, an unbinned event distribution may not even exist -- for instance
if the $x$ and $y$ values correspond to discrete PHA (spectral energy) channels.
Also, {\em some}\/ binning can be desirable, for instance, in cases where the
dynamic range of the data under consideration is large. If the bin size is
sufficiently small, the unavoidable loss of spatial resolution introduced by
binning the raw event distribution may be a small price to be paid for a data
array of manageable size.

Smoothing of high-resolution data is of interest whenever the signal (defined as
the number of counts per pixel above the expected background) in the region of
interest in $x-y$ space is low, i.e.\ is less than or of the order of 10, after
the raw event distribution has been sorted into intervals whose size matches
approximately or exceeds the instrumental resolution. It is crucial in this
context that the observed count statistics are not taken at face value but are
corrected for background, which may be internal, i.e.\ originating from the
detector (more general: the instrumental setup), or external. If this correction
is not applied, the observed intensity (counts) distribution $I(x,y)$ may be
high across the region of interest, suggesting good count statistics, even if
the signal above the background that we are interested in is actually low and
poorly sampled. The statistics of the observed counts alone can thus be a poor
indicator of the need for image smoothing.

Rebinning the data set into larger, and thus fewer, intervals improves the count
statistics per pixel and reduces the need for smoothing.  This is also the basic
idea behind smoothing with a kernel of the form
\begin{equation}
  {\cal K}(\vec{r}\,',\sigma) = \left\{ 
  \begin{array}{ll}  1/(\pi \sigma^2) & \mbox{where \,$|\vec{r}\,'| < \sigma $} \\
			     0   & \mbox{elsewhere}  
	\end{array} \right. 
  \label{tophat}
\end{equation} 
(circular `top-hat' or `box-car' filter of radial size $\sigma$), the only
difference being that smoothing occurs semi-continuously (the step size being
given by the bin size of the original data) whereas rebinning requires an
additional phase information [the offset of the boundaries of the first bin with
  respect to some point of reference such as the origin of the $(x,y)$
  coordinate system]. However, when starting from an image binned at about the
instrumental resolution, both rebinning and conventional smoothing share a well
known drawback, namely that any improvement in the count statistics occurs at
the expense of spatial resolution.

\section{Adaptive Kernel Smoothing} 

Although conventional smoothing algorithms usually employ more sophisticated
functional forms for the kernel than the above `top-hat' filter (the most
popular probably being a Gaussian), the problem remains that a kernel of fixed
size is ill-suited for images that feature real structure on various scales,
some of which may be much smaller or much larger than the kernel size. In such a
situation, small-scale features tend to get over-smoothed while large-scale
structure remains under-smoothed. {\em Adaptive-kernel smoothing} (AKS) is the
generic term for an approach developed to overcome this intrinsic limitation by
allowing the kernel to vary over the image and adopt a position-dependent
`natural' size.

AKS is closely related to the problem of finding the optimal adaptive kernel
estimator of the probability density distribution underlying a measured,
unbinned event distribution. The advantages of adaptive kernel estimators for
the analysis of discrete, and in particular one-dimensional, astronomical data
have been discussed by various authors (e.g.\ Thompson 1990; Pisani 1993, 1996;
Merritt and Tremblay 1994; Vio et al.\ 1994). An overview of adaptive filtering
techniques in two dimensions is given by Lorenz et al.\ (1993).

A common feature of all non-parametric adaptive kernel algorithms is that the
`natural' smoothing scale for any given position is determined from the number
of counts accumulated in its immediate environment. Following the aforementioned
principle, smoothing occurs over a large scale where few counts have been
recorded, and over a small scale where count statistics are good. AKS algorithms
differ, however, in the prescription that defines how the amplitude of the local
signal is to be translated into a smoothing scale.

A criterion widely used for discrete data is that of Silverman (1986).  It
determines the size, $\sigma$, of the local kernels relative to that of some
global (i.e.\ non-adaptive, fixed) kernel ($\sigma_{\mbox{\scriptsize const}}$)
by introducing a scaling factor which is the inverse square root of the ratio of
the globally smoothed data to their logarithmic mean.  For images, and using the
same notation as before, this means
\begin{equation}
   \sigma(\vec{r}\,) = \sqrt{\frac{\langle 
                          I'_{\mbox{\scriptsize const}}(\vec{r}\,)
                          \rangle_{\mbox{\scriptsize log}}}
                          {I'_{\mbox{\scriptsize const}}(\vec{r}\,)}},
\label{sman}
\end{equation}
where $\log_{10} \langle I'_{\mbox{\scriptsize
    const}}(\vec{r}\,)\rangle_{\mbox{\scriptsize log}} = \langle \log_{10}
I'_{\mbox{\scriptsize const}}(\vec{r}\,)\rangle $, and $I'_{\mbox{\scriptsize
    const}}(\vec{r}\,)$ represents the convolution of the measured data with a
kernel of fixed size $\sigma_{\mbox{\scriptsize const}}$.  However, whether or
not this approach yields satisfactory results depends strongly on the choice of
the global smoothing scale $\sigma_{\mbox{\scriptsize const}}$ (Vio et
al.\ 1994). In the context of discrete data sets, Pisani (1993) suggested a
least-squares cross-validation procedure to determine an optimal global kernel
size in an iterative loop. However, for binned data covering a large dynamical
range (see Section~\ref{example} for an example), the dependence of the result
on the size of the global kernel becomes very sensitive indeed, and the
iteration becomes very time-consuming.  Also the dependence on the somewhat
arbitrary scaling law (eq.~\ref{sman}) remains. Other adaptive filtering
techniques discussed recently in the literature include the {\sc hfilter}
algorithm for square images (Richter et al.\ 1991, see also Lorenz et al.\ 1993)
and the {\sc akis} algorithm of Huang \& Sarazin (1996).  Closely related are
image decomposition techniques including wavelet-based algorithms (Starck \&
Pierre 1998, and references therein) and adaptive binning (e.g., Sanders \&
Fabian 2001, Cappellari \& Copin 2003, Diehl \& Statler 2005).

In the following, we present {\sc asmooth}, an AKS algorithm for images,
i.e.\ binned, two-dimensional datasets of any size, which determines the local
smoothing scale from the requirement that the {\em above the
  background-corrected}\/ signal-ro-noise ratio (s.n.r.) of any signal enclosed
by the kernel must exceed a certain, preset value. The algorithm is similar to
{\sc akis} (Huang \& Sarazin 1996) in that it employs a signal-to-noise ratio
(s.n.r.) criterion to determine the smoothing scale\footnote{We stress that {\sc
    asmooth} was developed completely independently.}. However, other than {\sc
  akis}, {\sc asmooth} does not require any initial fixed-kernel smoothing but
determines the size of the adaptive kernel directly and unambiguously from the
unsmoothed input data. {\sc Asmooth} also goes beyond existing AKS algorithms in
that its s.n.r.\ criterion takes the background (instrumental or other) of the
raw image into account. This leads to significantly improved noise suppression
in the case of large-scale features embedded in high background. Our approach
yields smoothed images which feature a near-constant (or, alternatively,
minimal) signal-to-noise ratio above the background in all pixels containing a
sufficient number of counts. In contrast to most other algorithms which require
threshold values to be set (e.g., for the H coefficients in the case of the {\sc
  hfilter} technique), {\sc asmooth} is intrinsically non-parametric.  The only
external parameters that need to be specified are the minimal and, optionally,
maximal signal-to-noise ratios (above the background) required under the kernel.

The simplicity of the determination of the local smoothing scale from
the counts under the kernel and an estimate of the local background
greatly facilitates the translation of the smoothing prescription into
a simple and robust computer algorithm, and also allows a
straightforward interpretation of the resulting smoothed image.

\section{Description of the algorithm}
\label{descr}

{\sc Asmooth} adjusts the smoothing scale (i.e.\ the size of the smoothing
kernel) such that, at every position in the image, the resulting smoothed data
values share the same signal-to-noise ratio with respect to the background; one
may call this the `uniform significance' approach. The only external parameter
required by {\sc asmooth} is the desired minimal s.n.r.,
$\tau_{\mbox{\scriptsize min}}$.

In order to ensure that statistically significant structure is not over-smoothed
to an s.n.r.\ level much higher than $\tau_{\mbox{\scriptsize min}}$, an
s.n.r.\ range can be specified as a pair of $\tau_{\mbox{\scriptsize min}}$,
$\tau_{\mbox{\scriptsize max}}$ values. Note, however, that the maximal
s.n.r.\ criterion is a soft one and, also, is applied only at scales larger than
the instrumental resolution (which is assumed to be similar to, or larger than,
the pixel scale); under no circumstances will {\sc asmooth} blur significant
pointlike features (pixels whose s.n.r.\ in the unsmoothed image exceeds
$\tau_{\mbox{\scriptsize min}}$) in order to bring their s.n.r.\ value down below
the $\tau_{\mbox{\scriptsize max}}$ threshold.

The background-corrected s.n.r.\ of any features is computed according to one of
two definitions. The weaker requirement is given by the definition of the
``significance of detection'' above the background,
\begin{equation}
    \tau = \frac{(N_{\rm src}-N_{\rm bkg})}{\Delta N_{\rm bkg}},
\end{equation}
where $N_{\rm src}$ and $N_{\rm bkg}$ are the number of counts under the
smoothing kernel and the background kernel, respectively, and $\Delta N_{\rm
  bkg}$ is the $1\sigma$ uncertainty of the background counts. Alternatively,
and by default, the more stringent definition of the ``significance of the
source strength measurement'' can be used,
\begin{equation}
    \tau = \frac{(N_{\rm src}-N_{\rm bkg})}{\sqrt{(\Delta N_{\rm
    src})^2+(\Delta N_{\rm bkg})^2}},
\end{equation}
with $\Delta N_{\rm src}$ being the $1\sigma$ uncertainty of the source counts
accumulated under the smoothing kernel. Either of these definitions assume
Gaussian statistics by implying that the $n\sigma$ error of a measurement is
equal to $n$ times the $1\sigma$ error.

In addition to the desired s.n.r.\ limits $\tau_{\mbox{\scriptsize min, max}}$,
estimates of the background $I_{\mbox{\scriptsize bkg}}$ and the associated
background error $\Delta I_{\mbox{\scriptsize bkg}}$ are optional additional
parameter. To allow background variations across the image to be taken into
account, $I_{\mbox{\scriptsize bkg}}$ and $\Delta I_{\mbox{\scriptsize bkg}}$
can be supplied as images of the same dimensions as the raw image; in the case
of a flat background $I_{\mbox{\scriptsize bkg}}$ and $\Delta
I_{\mbox{\scriptsize bkg}}$ reduce to global estimates of the background and
background error per pixel, i.e., single numbers. Note that, more often than
not, $\Delta I_{\mbox{\scriptsize bkg}} \neq \sqrt{I_{\mbox{\scriptsize bkg}}}$
as the background estimate will originate from model predictions rather than
being the result of another counting experiment. If no background information is
supplied, {\sc asmooth} determines a local background from an annular region
around the adaptive smoothing kernel, extending from 3-4$\sigma$ for a Gaussian
kernel, and from 1-4/3$\sigma$ for a top-hat kernel.

Internally, the threshold s.n.r.\ values $\tau_{\mbox{\scriptsize min}}$,
$\tau_{\mbox{\scriptsize max}}$ are translated into a minimal and a maximal
integral number of counts, $N_{\mbox{\scriptsize min}}$, $N_{\mbox{\scriptsize
    max}}$, to be covered by the kernel. More precisely, the criterion is that
\begin{equation}
    N_{\mbox{\scriptsize min}} \leq 
    I'(\vec{r}\,)/{\cal K}(\vec{0},\sigma(\vec{r}\,)) \la
                   N_{\mbox{\scriptsize max}}
 \label{mc_crit}
\end{equation}
where $\sigma(\vec{r}\,)$ is the characteristic, position-dependent scale of the
respective kernel. $N_{\mbox{\scriptsize min,max}}$ in eq.~\ref{mc_crit} are
determined from the definition of the minimal and maximal s.n.r. value
$\tau_{\mbox{\scriptsize min,max}}$,
\begin{equation}
  \tau_{\mbox{\scriptsize min,max}} = \frac{N_{\mbox{\scriptsize min,max}}
               -N_{\mbox{\scriptsize bkg}}}
         {\sqrt{N_{\mbox{\scriptsize min,max}}+\Delta N^2_{\mbox{\scriptsize 
                bkg}}}},
  \label{taudef}
\end{equation}
where, in analogy to the definition of $N_{\mbox{\scriptsize min,max}}$
(cf. eqs.~\ref{convol},\ref{mc_crit}), $N_{\mbox{\scriptsize bkg}}$ and $\Delta
N_{\mbox{\scriptsize bkg}}$ are the integral number of background counts under
the respective kernel and the associated error. From eq.~\ref{taudef} follows
\begin{eqnarray}
    N_{\mbox{\scriptsize min,max}} = 
    N_{\mbox{\scriptsize bkg}} & + & \frac{1}{2}\,
                         \tau_{\mbox{\scriptsize min,max}}^2 
    \nonumber \\
                         & + & \tau_{\mbox{\scriptsize min,max}} \,
                         \sqrt{N_{\mbox{\scriptsize bkg}} + 
                               \Delta N^2_{\mbox{\scriptsize bkg}} +
                         \frac{1}{4}\, \tau_{\mbox{\scriptsize min,max}}^2}.
 \label{n_min}
\end{eqnarray}
For an adaptive circular top-hat kernel of size $\sigma(\vec{r}\,)$
(cf.\ eq.~\ref{tophat}), eq.~\ref{mc_crit} translates into $N_{\mbox{\scriptsize
    min}} \leq \pi\,\sigma (\vec{r}\,)^2 \,I' (\vec{r}\,) \leq
N_{\mbox{\scriptsize max}}$, and the interpretation is straightforward: at least
$N_{\mbox{\scriptsize min}}$, but no more than $N_{\mbox{\scriptsize max}}$,
counts are required to lie within the area $\pi\,\sigma (\vec{r}\,)^2$ that the
smoothing occurs over. In the case of a uniform background, the value of
$N{\mbox{\scriptsize bkg}}$ in eq.~\ref{n_min} is simply given by
$n_{\mbox{\scriptsize bkg}}\pi\,\sigma (\vec{r}\,)^2$ where
$n_{\mbox{\scriptsize bkg}}$ is the global background level per pixel in the
input image.

For any given pair of $(N_{\mbox{\scriptsize min}},N_{\mbox{\scriptsize max}})$
values, a Gaussian kernel
\begin{equation}
  {\cal K}(\vec{r}-\vec{r}\,',\sigma(\vec{r}\,)) = \frac{1}{2\,\pi\,\sigma(\vec{r}\,)^2} \;
		\exp \left( - \frac{|\vec{r}-\vec{r}\,'\,|^2}{2\sigma(\vec{r}\,)^2} \right)
  \label{gaussian}
\end{equation} 
will yield considerably larger effective smoothing scales than a top-hat, as, in
two dimensions, more than 60 per cent of the integral weight fall outside a $1\,
\sigma$ radius, whereas, in the case of a circular top-hat kernel, all of
$N_{\mbox{\scriptsize min}}$ needs to be accumulated within a $1\, \sigma$
radius. (Note that, according to Eq.~\ref{gaussian}, it is the weights {\em per
  unit area} that follow a Gaussian distribution. The weights per radial annulus
{\em do not}, which is why, for the kernel defined in Eq.~\ref{gaussian}, the
fraction of the integral weight that falls outside the $1\, \sigma$ radius is
much larger than the 32 per cent found for a one-dimensional Gaussian.) Which
kernel to use is up to the user: {\sc asmooth} offers a choice of Gaussian
(default) and circular top-hat.

The algorithm is coded such that the adaptively smoothed image is accumulated in
discrete steps as the smoothing scale increases gradually, i.e.
\begin{equation}
    I'_{\mbox{\scriptsize AKS}} (\vec{r}\,) = \sum_i I'_i(\vec{r}\,) 
                   =  \sum_i I_i(\vec{r}\,) \otimes {\cal K}(\vec{r},\sigma_i) \;,
	\label{AKS_acc}
\end{equation}
where $\sigma_i$ starts from an initial value $\sigma_0$ which is matched to the
intrinsic resolution of the raw image (i.e., the pixel size), and
$I_i(\vec{r}\,)$ is given by
\begin{equation}
   I_i(\vec{r}\,) = \left\{ \begin{array}{ll}  
	I(\vec{r}\,) & \mbox{where \,$N_{\mbox{\scriptsize min}} \leq
                             I'(\vec{r}\,)/{\cal K}(\vec{0},
                             \sigma_i) \leq N_{\mbox{\scriptsize max}}$}\\
                     & \mbox{\quad and \,$I(\vec{r}\,)\not\in I_j(\vec{r}\,),
                       j<i$} \label{maskcrit}  \\
		 0   & \mbox{elsewhere.}  
	                    \end{array} \right. 
\end{equation}
The adaptively smoothed image is thus accumulated in a ``top-down'' fashion with
respect to the observed intensities as {\sc asmooth} starts at small kernel
sizes to smooth the vicinity of the brightest pixels, and then increases the
kernel size until, eventually, only background pixels contribute. Note that
condition \ref{maskcrit} ensures that pixels found to contain sufficient signal
at a scale $\sigma_i$ will not contribute to the image layers $I'_j, (j>i)$
subsequently produced with smoothing scales $\sigma_j>\sigma_i$.  Consequently,
each feature is smoothed at the smallest scale at which it reaches the required
background-corrected s.n.r.\ (see eq.~\ref{mc_crit}), and low-s.n.r.\ regions
are smoothed at an appropriately large scale even in the immediate vicinity of
image areas with very high s.n.r.

In order to take full advantage of the resolution of the unbinned image, the
size $\sigma_0$ of the smallest kernel is chosen such that the area enclosed by
${\cal K}(\vec{r},\sigma_0)$ is about one pixel.  For the circular top-hat
filter of eq.~\ref{tophat} this means $\sigma_0 = 1/\sqrt{\pi}$; for the
Gaussian kernel of eq.~\ref{gaussian} we have $\sigma_0 =
1/\sqrt{9\pi}$. Subsequent values of $\sigma_i \,(i>0)$ are determined from the
requirement that eq.~\ref{mc_crit} be true. If a near-constant s.n.r. value is
aimed at with high accuracy, i.e., if a $\tau_{\mbox{\scriptsize max}}$ value
very close to $\tau_{\mbox{\scriptsize min}}$ is chosen, the smoothing scale
$\sigma_i$ will grow in very small increments, and the smoothing will proceed
only slowly. In all our applications we found values of $\tau_{\mbox{\scriptsize
    max}} \ga 1.1\times \tau_{\mbox{\scriptsize min}}$ to yield a good
compromise between CPU time considerations and a sufficiently constant
signal-to-noise ratio of the smoothed image. If no value for the optional input
parameter $\tau_{\mbox{\scriptsize min}}$ is supplied by the user, the code
therefore assumes a default value of $\tau_{\mbox{\scriptsize max}} =
\tau_{\mbox{\scriptsize min}}+1$ which meets the above requirement for all
reasonable values of $\tau_{\mbox{\scriptsize min}}$.

While the intrinsic resolution of the raw image (i.e. the pixel size) determines
the smallest kernel size $\sigma_0$, the size of the image as a whole represents
an upper limit to the size of the kernel.  Although the convolution can be
carried out until the numerical array representing the kernel is as large as the
image itself, this process becomes very CPU time intensive as $\sigma_i$
increases. Once the smoothing scale has exceeded that of the largest structure
in the image, the criterion of eq.~\ref{mc_crit} can never be met as only
background pixels contribute. Since the only features left unsmoothed at this
stage are insignificant background fluctuations, the algorithm then smoothes the
remaining pixels with the largest possible kernel. Unavoidably, the
signal-to-noise ratio of these last background pixels to be smoothed does not
meet the condition of eq.~\ref{mc_crit}. Note that in the case of the background
being determined locally from the data themselves (the default), it is the size
of the background kernel (an annulus around the smoothing kernel) that reaches
the limit first. Hence, for a Gaussian, the largest smoothing scale (1$\sigma$
size of the kernel) is 1/8 of the image size, for a top-hat it is 3/8 of the
image size.

Since the algorithm uses fast-fourier transformation (FFT) to compute the many
convolutions required, the overall performance of the code is significantly
improved if the image size in pixels is an integer power of two. The smoothed
image obtained from the above procedure strictly conserves total counts (within
the limitations set by the computational accuracy) and provides a fair
representation of the original data at all positions.

\section{Performance of the algorithm}

\label{example}
We first demonstrate the performance of our algorithm by applying it to an image
of X-ray emission from a massive cluster of galaxies taken with the ACIS-S
detector on board the Chandra X-ray Observatory. Then we use simulated data to
test how faithfully {\sc Asmooth} reproduces the true count distribution of the
input image.

\subsection{Results for Chandra ACIS-S data}

Because of the large range of scales at which features are detected in the
selected Chandra observation, this X-ray image is ideally suited for a
demonstration of the advantages of AKS. If photon noise is to be suppressed
efficiently, the very extended emission from the gaseous intracluster medium
needs to be smoothed at a rather large scale. At the same time, a small
smoothing scale, or no smoothing at all, is required in high-s.n.r.\ regions in
order to retain the spatial resolution in the vicinity of bright point-like
sources (stars, QSOs, AGN) superimposed on the diffuse cluster
emission. Fixed-kernel smoothing can meet, at most, one of these requirements at
a time.

Our choice of data set has the additional advantage that the selected image was
taken with an X-ray detector that is also very sensitive to charged particles
which makes the image particularly well suited to emphasize the importance of a
proper treatment of the background.

\begin{figure*}
	 \parbox{0.33\textwidth}{
	 \epsfxsize=0.32\textwidth
	 \epsffile{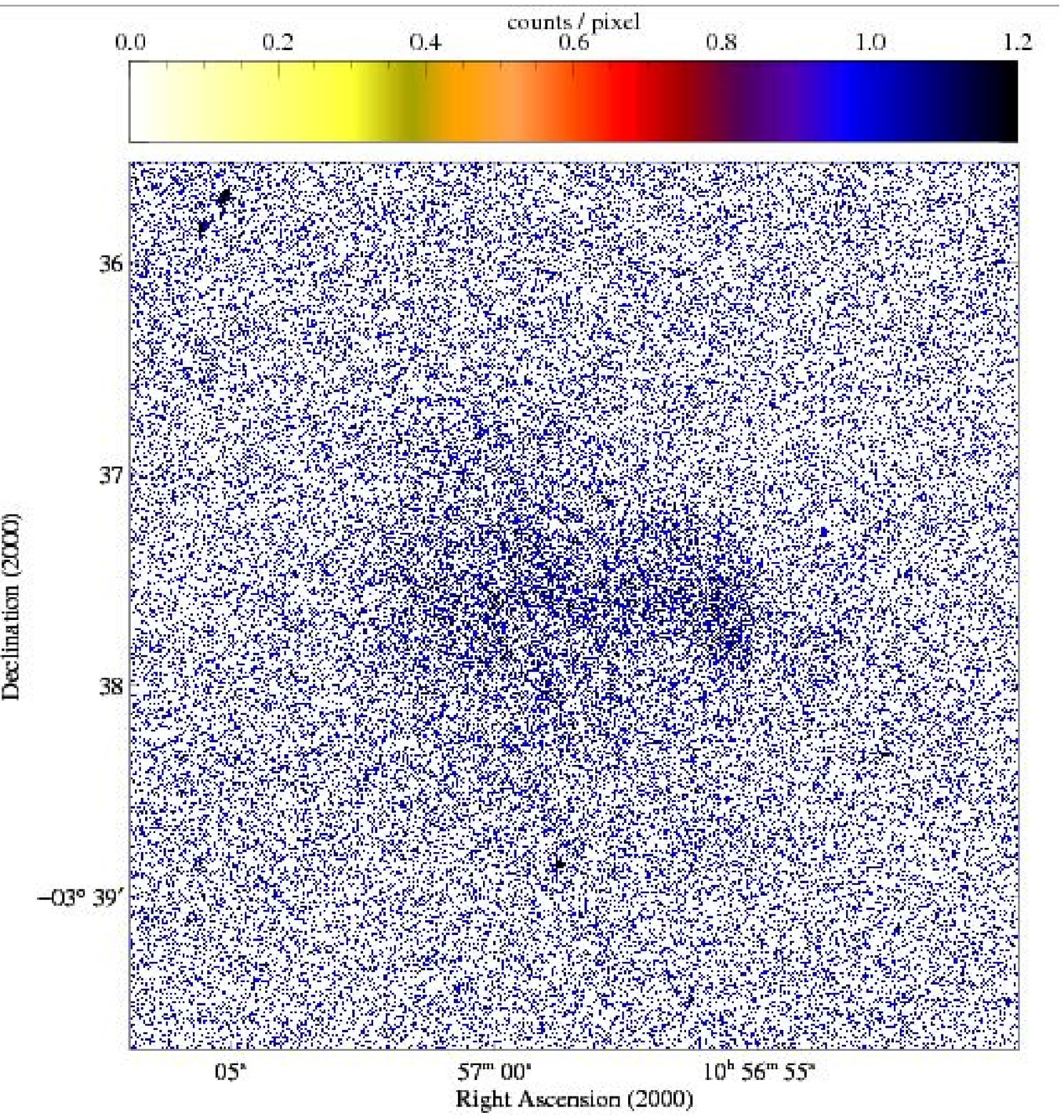}}\parbox{0.66\textwidth}{
\caption[]{{\bf Left}: X-ray emission detected with Chandra ACIS-S in the 0.5-7
  keV band in a $4.2\times 4.2$ arcmin$^2$ field around the cluster of galaxies
  MS\,1054.4$-$0321; the intensity scaling is logarithmic. Shown are the raw
  data. {\bf Below}: Results obtained with {\sc asmooth} (adaptive Gaussian
  kernel) for (top to bottom row) $\tau_{\mbox{\scriptsize min}} = 2$, 3, and
  4. In all cases $\tau_{\mbox{\scriptsize max}}$ was set to
  $\tau_{\mbox{\scriptsize min}} +1$. The three columns show (from left to
  right) the adaptively smoothed image (same intensity scaling as used for the
  image of the raw data shown in the single panel at the very top), a map of the
  kernel sizes ($1\sigma$ radius of a two-dimensional Gaussian) used by {\sc
    asmooth} in the smoothing of the raw data, and a map of the
  background-corrected s.n.r.\ of pixel values in the adaptively smoothed image.
  Note how {\sc asmooth} fully retains signal that reaches the specified
  background-corrected s.n.r.\ level while, at the same time, heavily
  suppressing background noise by applying a wide range of smoothing scales from
  much less than one to many tens of pixels. The majority of the pixels in the
  raw data contain insufficient signal to reach the specified s.n.r.\
  threshold at any permissible smoothing scale and are thus smoothed with the
  largest possible kernel. Note the correspondence between the outlines of
  regions with a signal-to-noise ratio of less than $\tau_{\mbox{\scriptsize
      min}}$ in the {\sc asmooth}ed images (right column) and the outlines of
  regions smoothed with largest possible kernel (center column). With the
  exception of a few very bright pixels whose s.n.r.\ exceeds
  $\tau_{\mbox{\scriptsize max}}$ in the raw data (i.e. without any smoothing)
  all other pixels in the {\sc asmooth}ed image contain signal whose
  background-corrected s.n.r.\ is near-constant at the specified level.
  \label{ms1054}
}}\\*[-7mm]
	 \parbox{0.33\textwidth}{
	 \epsfxsize=0.32\textwidth
	 \epsffile{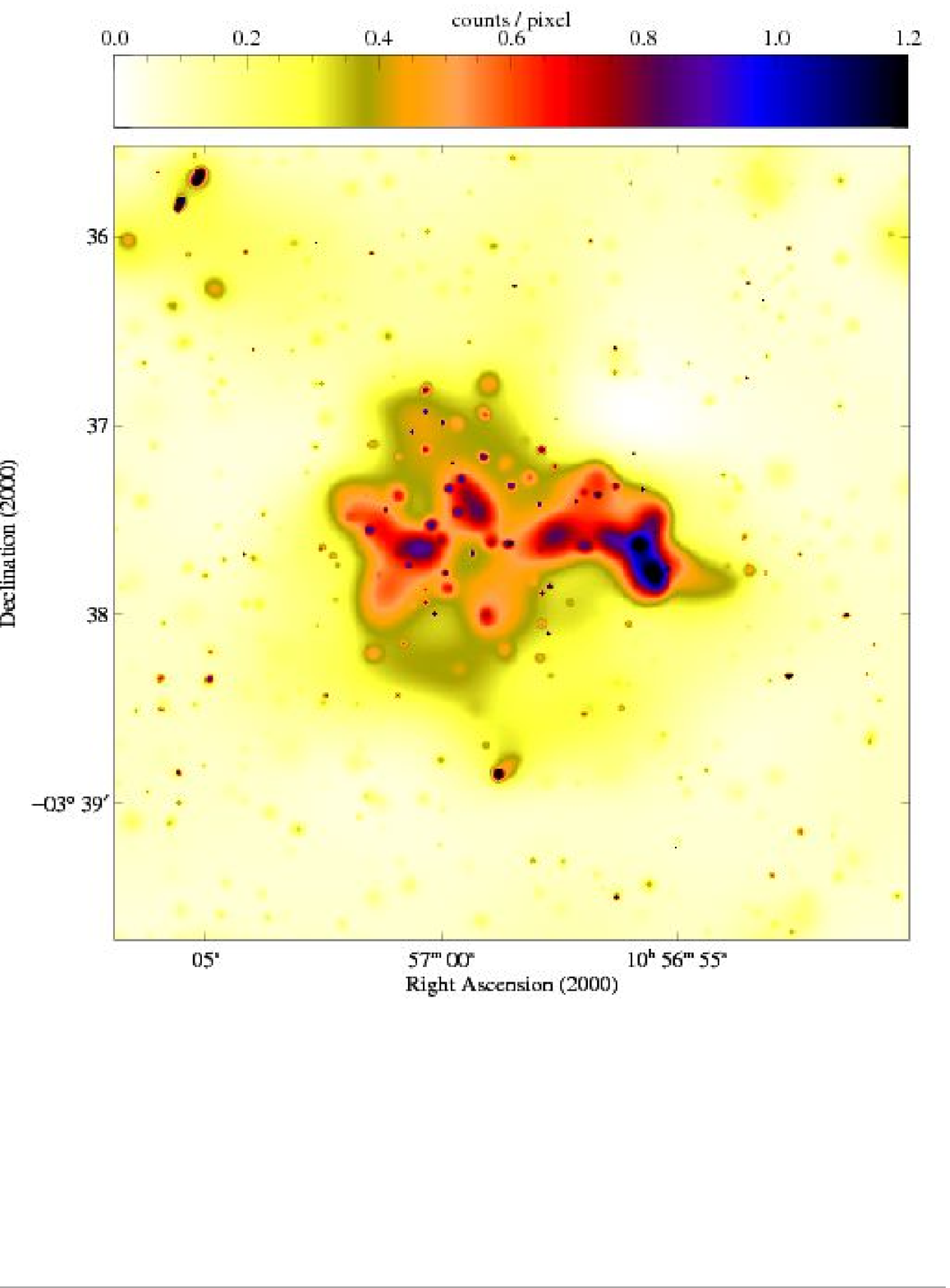}}
	 \parbox{0.33\textwidth}{
	 \epsfxsize=0.32\textwidth
	 \epsffile{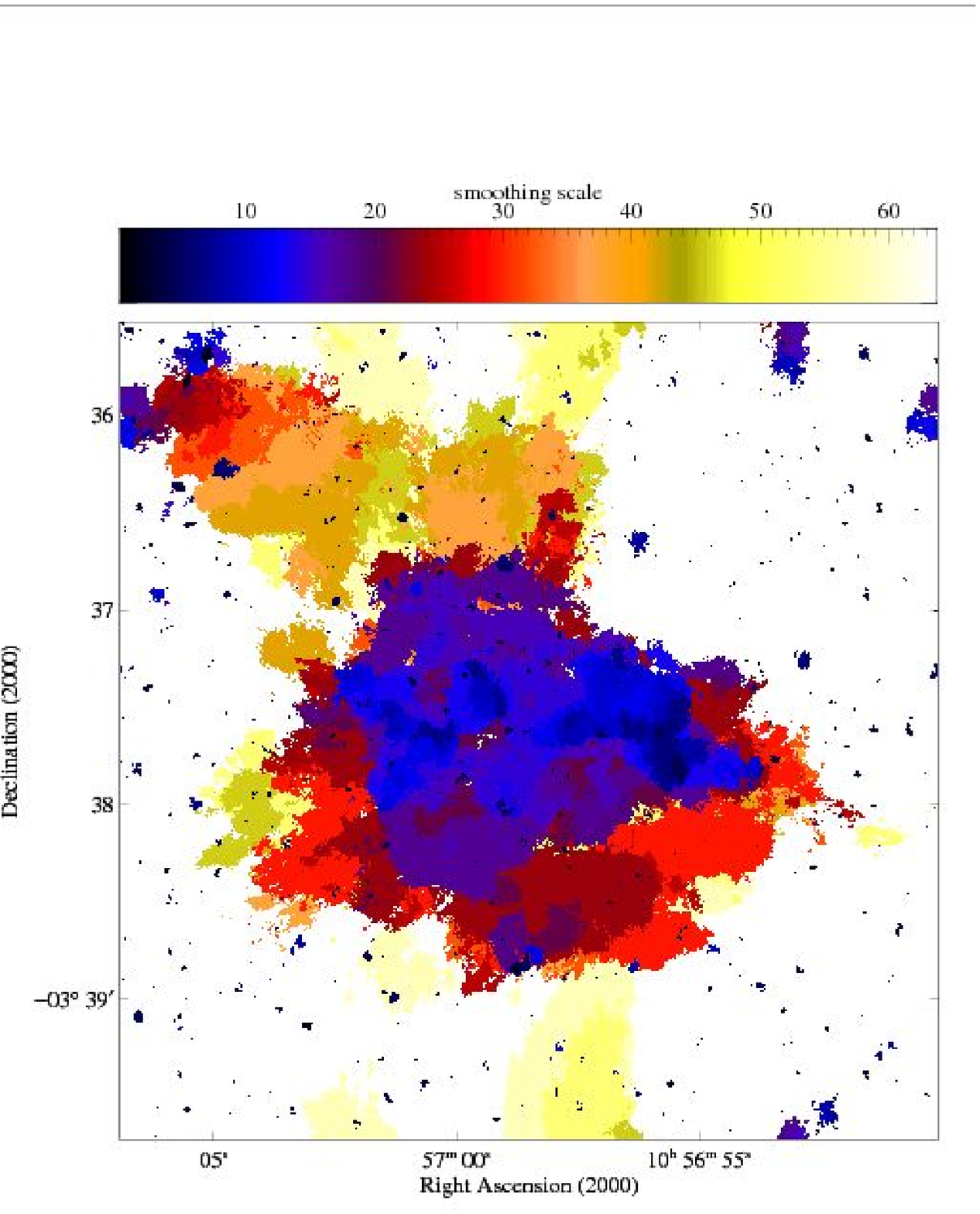}}
	 \parbox{0.33\textwidth}{
	 \epsfxsize=0.32\textwidth
	 \epsffile{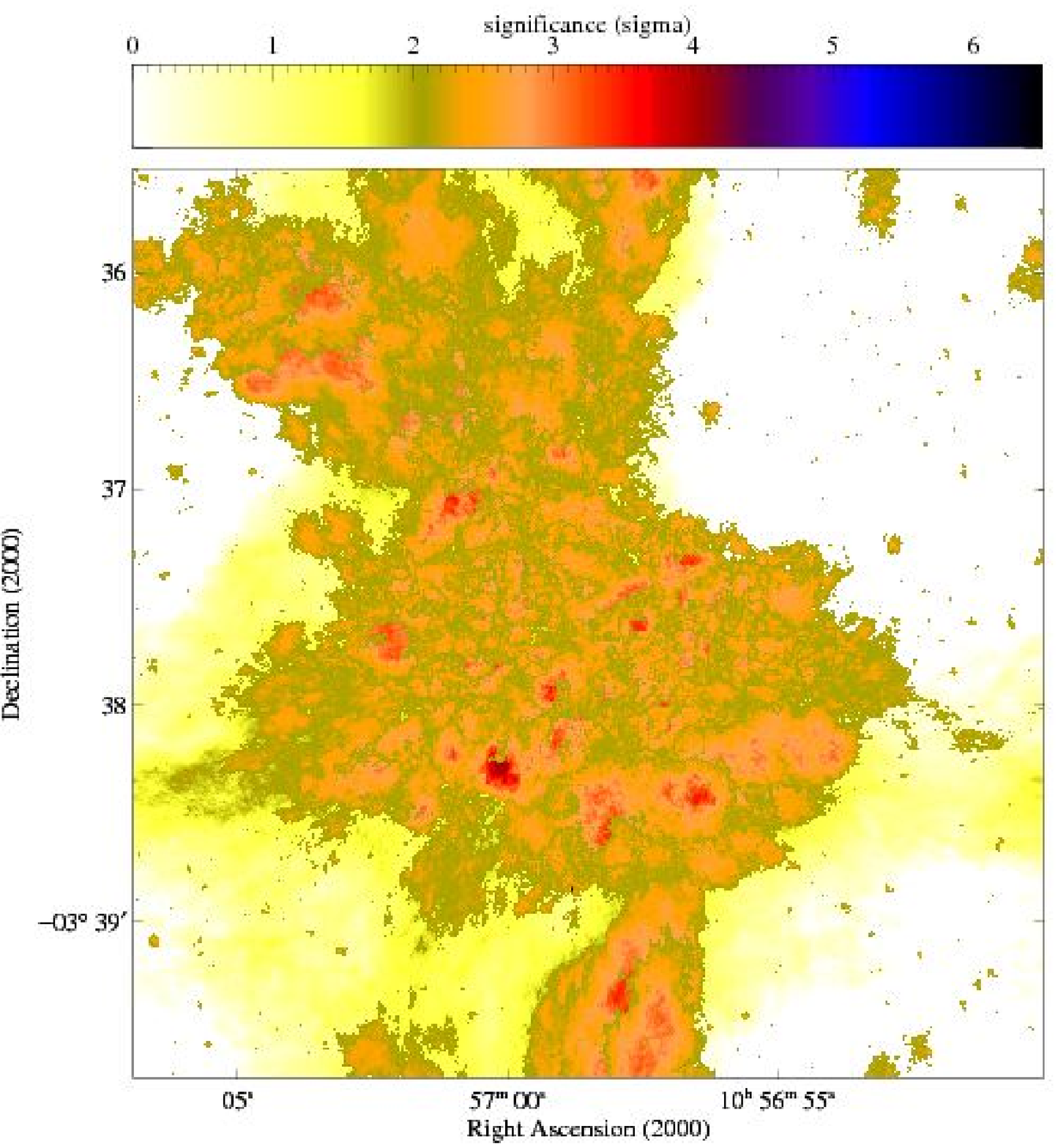}}\mbox{}\\*[-13mm]
	 \parbox{0.33\textwidth}{
	 \epsfxsize=0.32\textwidth
	 \epsffile{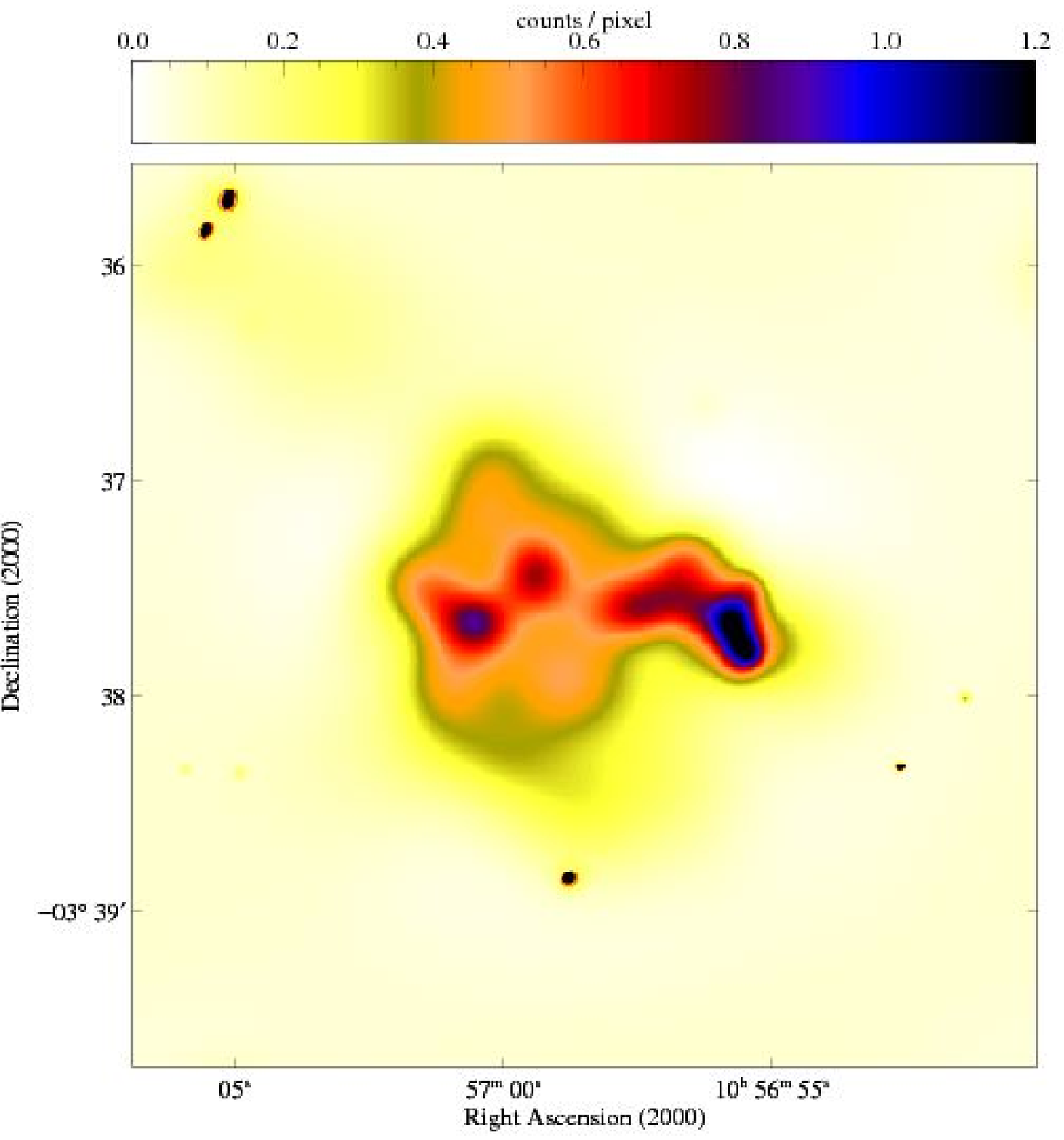}}
	 \parbox{0.33\textwidth}{
	 \epsfxsize=0.32\textwidth
	 \epsffile{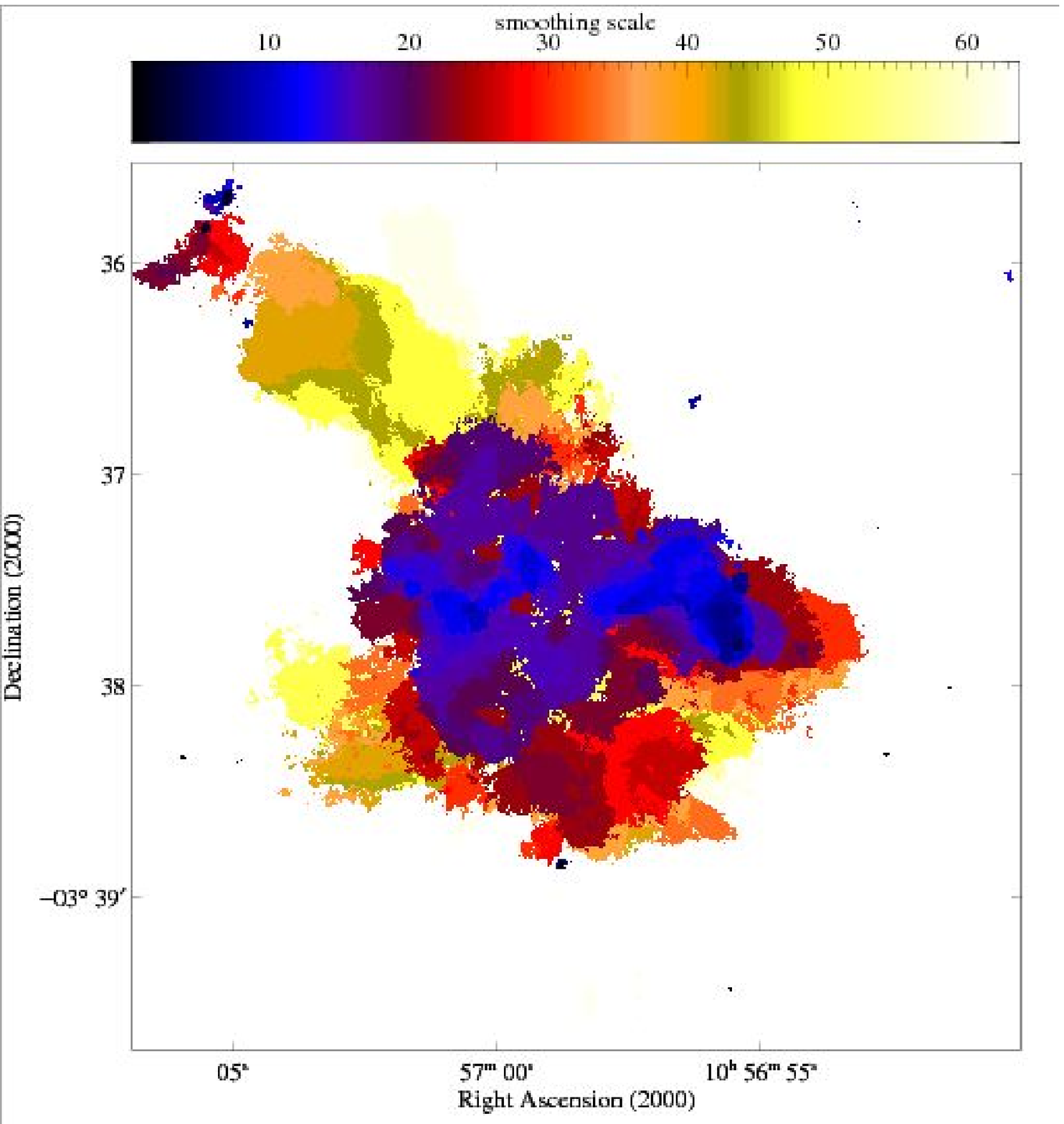}}
	 \parbox{0.33\textwidth}{
	 \epsfxsize=0.32\textwidth
	 \epsffile{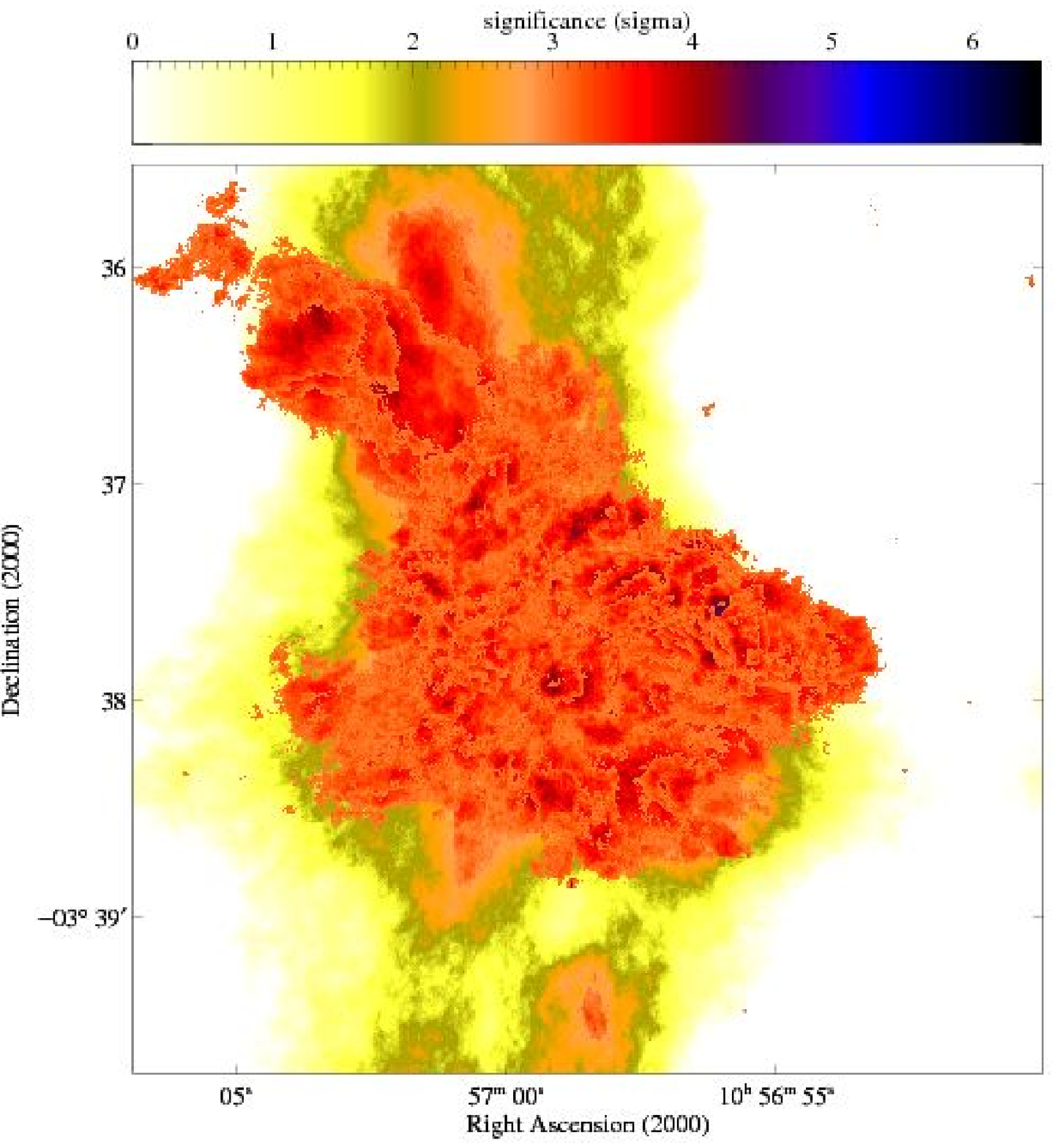}}\mbox{}\\*[-7mm]
	 \parbox{0.33\textwidth}{
	 \epsfxsize=0.32\textwidth
	 \epsffile{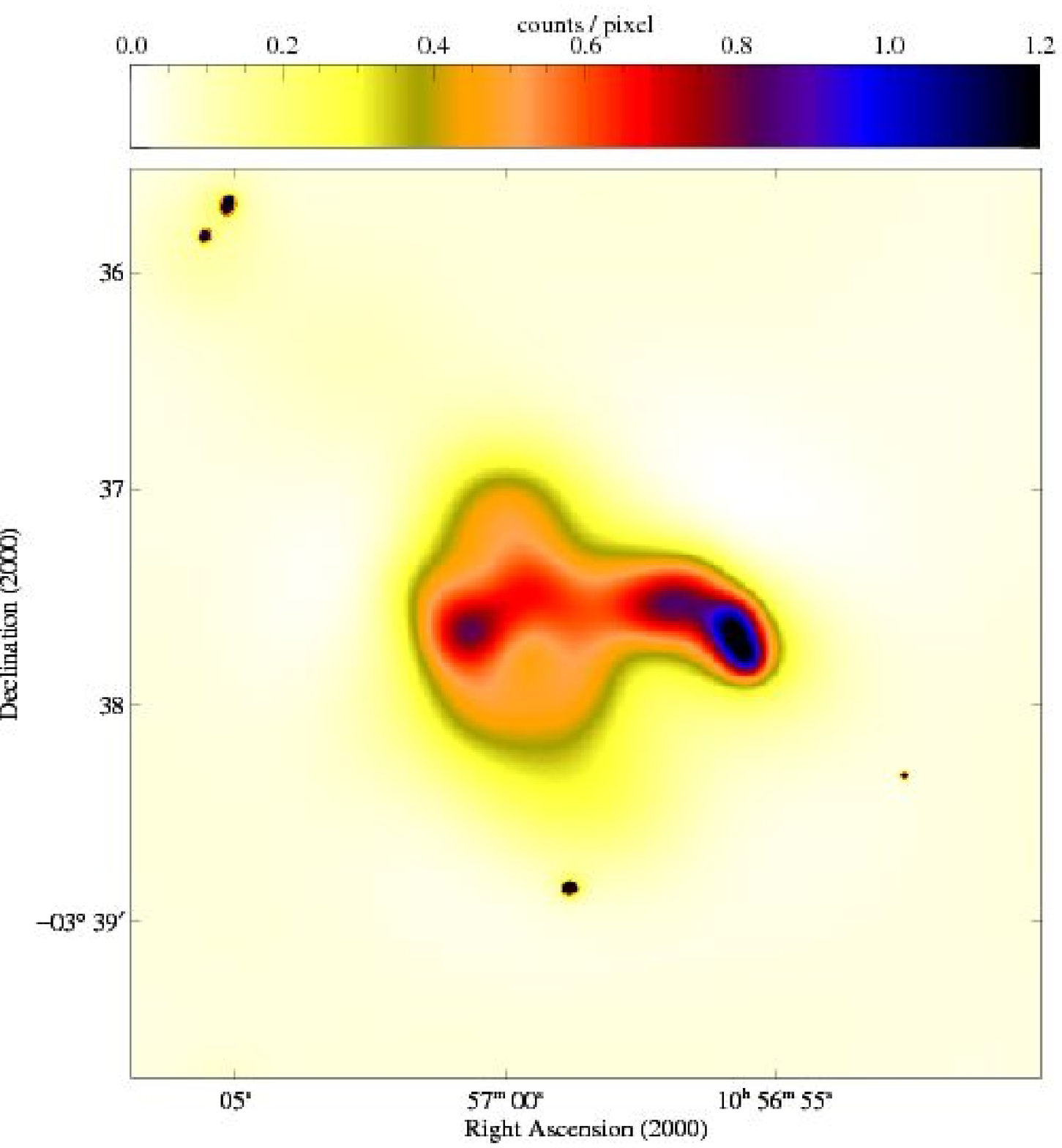}}
	 \parbox{0.33\textwidth}{
	 \epsfxsize=0.32\textwidth
	 \epsffile{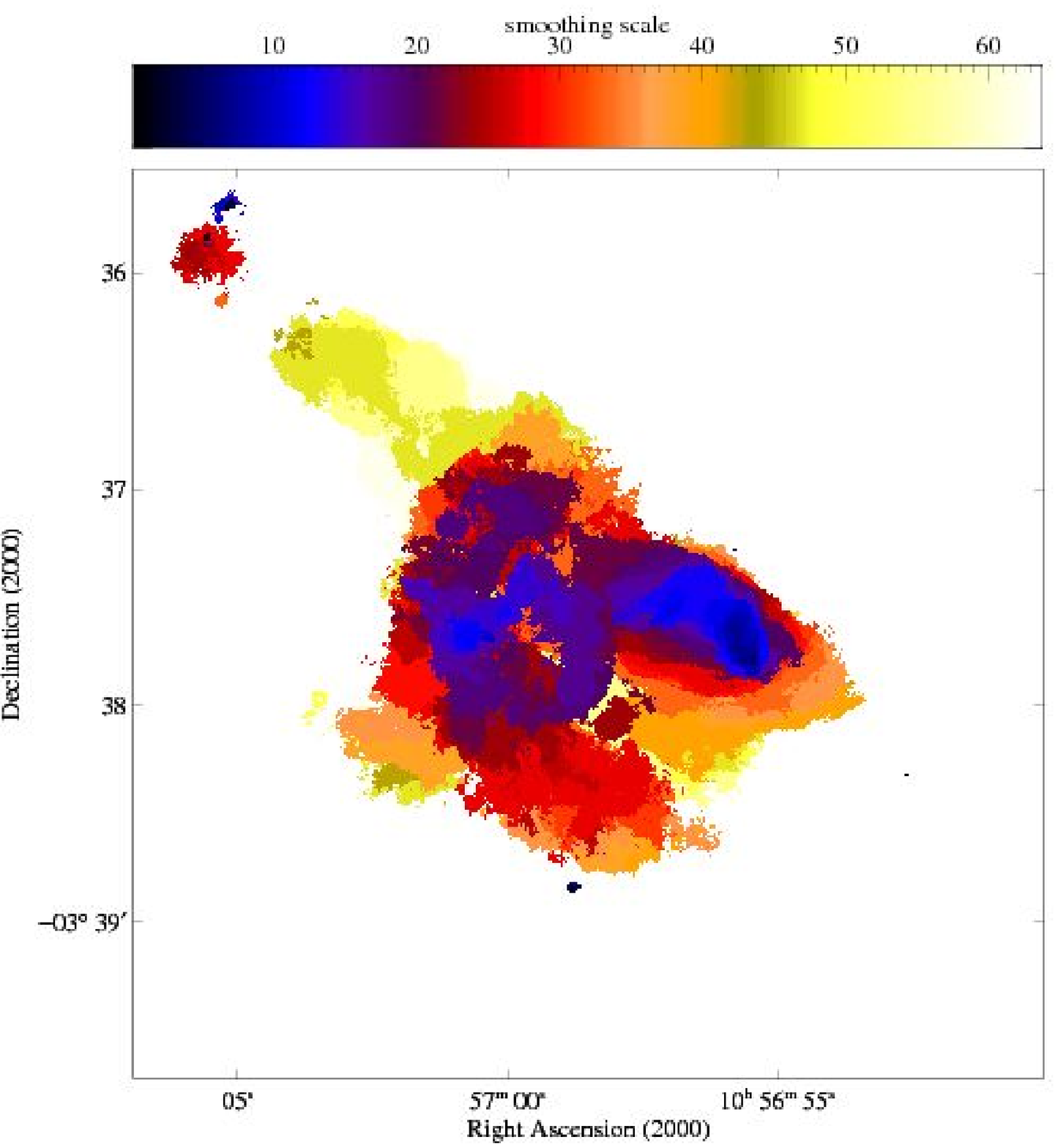}}
	 \parbox{0.33\textwidth}{
	 \epsfxsize=0.32\textwidth
	 \epsffile{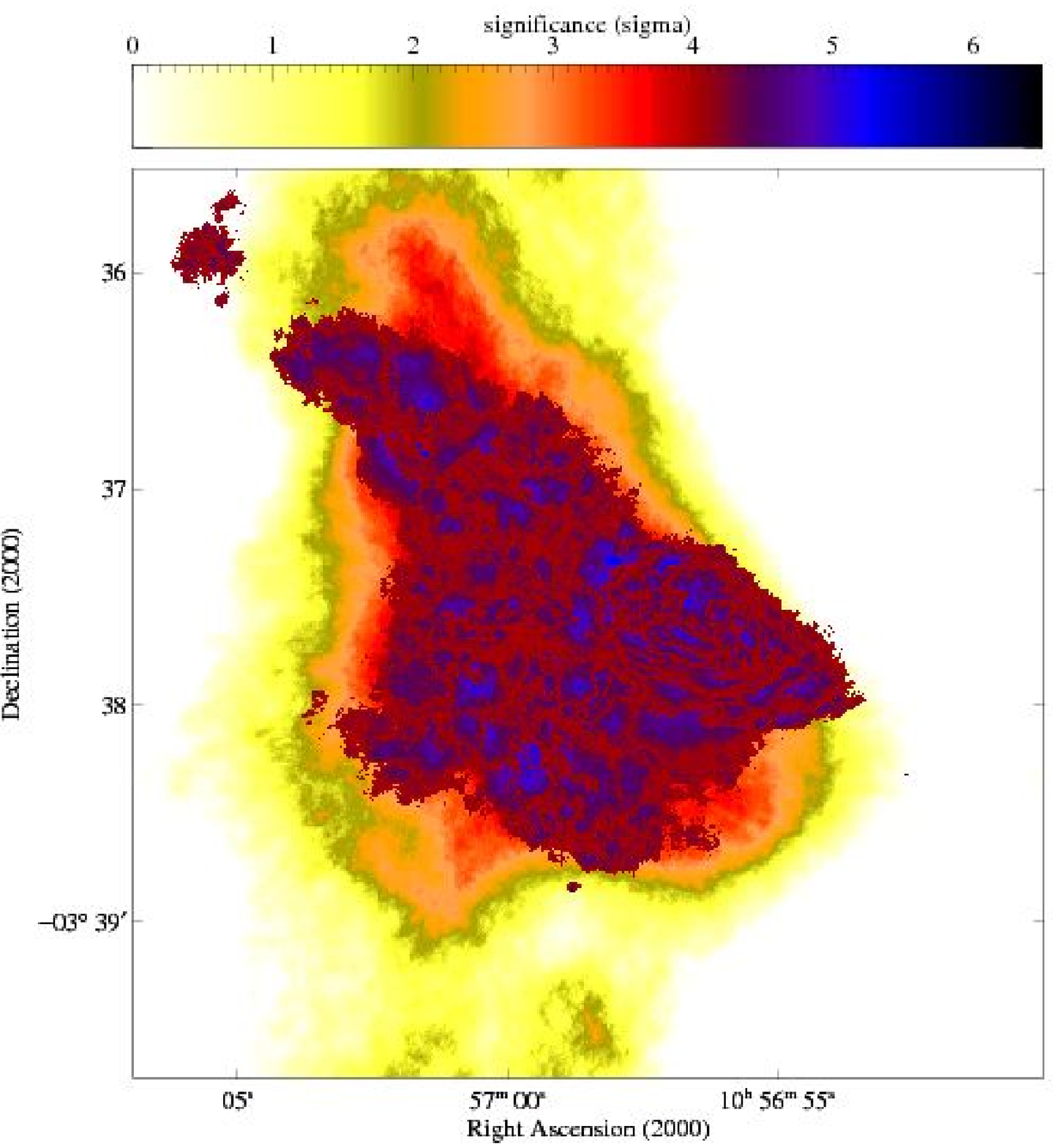}}
\end{figure*}

Fig.~\ref{ms1054} (top left panel) shows the raw counts detected with ACIS-S in
the 0.5 to 7 keV energy band in an 89 ks observation of the galaxy cluster
MS\,1054.4$-$0321. The diffuse emission originates from an electron-ion plasma
trapped in the gravitational potential well of the cluster and heated to
temperatures of typically $1\times 10^8$ K (corresponding to ${\rm k}T \approx
10$ keV). A detailed analysis of this observation is given by Jeltema et
al.\ (2001). The image shown here ($512\times 512$ pixel$^2$) covers a
subsection of the detector spanning a $4.2\times 4.2$ arcmin$^2$ region; the
pixel size of 0.492 arcsec corresponds roughly to one on-axis resolution element
of the telescope-detector configuration. Note the high background throughout the
image, caused by high-energy cosmic rays.

Fig.~\ref{ms1054} summarizes {\sc asmooth} results obtained with a Gaussian
kernel, and for s.n.r. target values of $\tau_{\mbox{\scriptsize
    min}}=2,\,3,\,4$ and $\tau_{\mbox{\scriptsize max}}=\tau_{\mbox{\scriptsize
    min}}+1$, in a three-by-three array of plots below the image of the raw
data.

\subsubsection{{\sc asmooth}ed images}

The left-most column of Fig.~\ref{ms1054} shows the adaptively smoothed images
for the three different $\tau_{\mbox{\scriptsize min}}$ values. {\sc Asmooth}
fully preserves the high information content of the raw data in the
high-s.n.r.\ regions corresponding to bright, small-scale features, while at the
same time heavily smoothing the low-s.n.r.\ regions of the image where the
signal approaches the background value. Note, however, that for small values of
$\tau_{\mbox{\scriptsize min}}$ (top row) the goodness of the local estimation
of the signal above the background is relatively low and noise is not removed
efficiently on all scales.

\begin{figure}
\epsfxsize=0.5\textwidth
\hspace*{-3mm}\epsffile{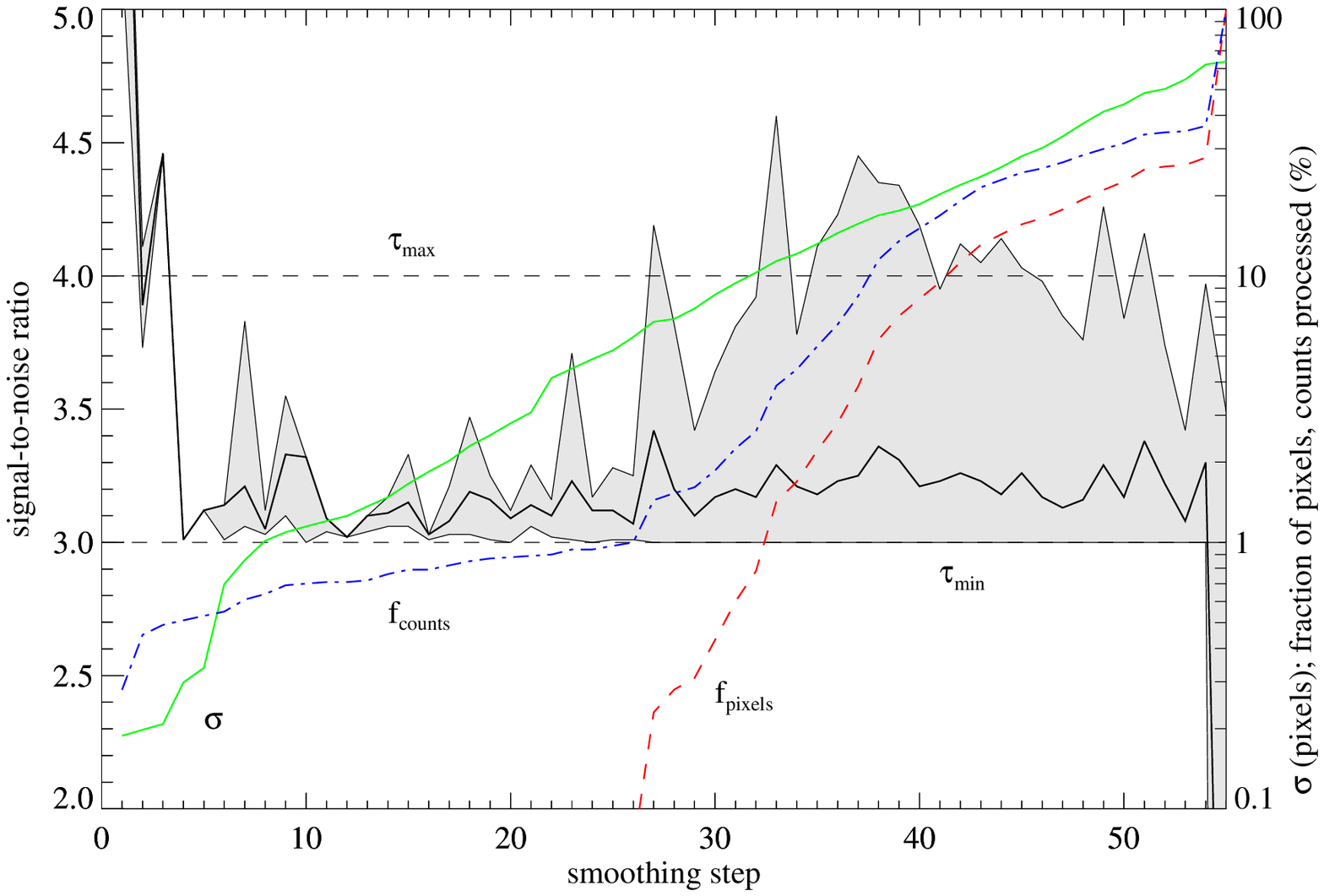}
\caption{{\sc Asmooth} processing parameters as a function of smoothing step for
  one of the examples shown in the Fig~\protect\ref{ms1054}
  ($\tau_{\mbox{\scriptsize min}}=3,\,\tau_{\mbox{\scriptsize max}}=4$). The
  shaded region and solid black line delineate the range of signal-to-noise
  ratios and their median value, respectively, for the set of image pixels
  processed in each step. The target values $\tau_{\mbox{\scriptsize min}}$ and
  $\tau_{\mbox{\scriptsize max}}$ are marked by the dashed horizontal lines.
  Note the nonlinear increase in the smoothing radius $\sigma$ (green line), the
  cumulative fraction of pixels processed (red line), and the cumulative
  fraction of counts processed. Note also that the criterion of
  eq.~\protect\ref{mc_crit} is a soft one as far as $N_{\rm max}$ ($\tau_{\rm
    max}$) is concerned: $\tau$ values greater than $\tau_{\rm max}$ are
  tolerated as long as the median value of the s.n.r.\ distribution for
  each smoothing step is smaller than $(\tau_{\rm max}+\tau_{\rm
    min})/2$. S.n.r.\ values exceeding $\tau_{\rm max}$ occur also at the
  smallest smoothing scales as the code is designed not to blur features whose
  s.n.r.\ is higher than $\tau_{\rm min}$ already in the raw data. }
\label{plotlog}
\end{figure}

Figure~\ref{plotlog} illustrates the gradual assembly of the adaptively smoothed
image by showing the change in various {\sc asmooth} parameters as a function of
smoothing step.

At the highest pixel intensities above the s.n.r.\ threshold, {\sc asmooth}
occasionally returns smoothed count values that {\em exceed}\/ the counts in the
unsmoothed image: the brightest point source in the raw data has a peak value of
126 counts, the {\sc asmooth}ed image features a value of 128.5 at the same
location. This is due to the fact that, although the corresponding pixels
themselves remain essentially unsmoothed and thus keep their original values,
there is an additional contribution from the larger sized kernels of
neighbouring pixels whose s.n.r.\ above the background falls short of
$\tau_{\mbox{\scriptsize min}}$. Since the final image is accumulated from the
partial images resulting from the successive convolution with the whole set of
differently sized kernels (cf.\ Eq.~\ref{AKS_acc}), the total smoothed intensity
can become larger than the actually observed counts at pixel positions where
$I(\vec{r}\,)\geq N_{\mbox{\scriptsize min}}$. This artifact is caused by the
limited resolution of the images and can be noticeable when a large bin size is
chosen for the original data. The integral number of counts in the image is
always conserved.

\subsubsection{Kernel size and s.n.r.\ maps}

As illustrated in Fig.~\ref{plotlog} {\sc asmooth} applies a wide range of
smoothing scales to the input data as the algorithm attempts to fulfill the
requirement given by Eq.~\ref{mc_crit} throughout the image. In addition to the
adaptively smoothed image, {\sc asmooth} also returns, in an IDL data structure,
maps of the background-corrected s.n.r.\ of the pixel values in the smoothed
image and the kernel size used in the smoothing process, respectively,

The second and third columns of the three-py-three panel in Fig.~\ref{ms1054}
show both of these maps for the {\sc asmooth} images shown in the first,
left-most column of plots. The maps of {\sc asmooth} kernel sizes as well as
Fig~\ref{plotlog} demonstrate how very small kernel sizes of less than or about
one pixel ($1\sigma$ radius) are assigned to very few bright pixels;
accordingly, these pixels remain essentially unsmoothed. For $\tau_{\rm min}=3$
(middle row), for instance, some 27 per cent of the image pixels are found to
satisfy the criterion of Eq.~\ref{mc_crit} at smoothing scales between one and
62 pixels (radius), while the majority of the remaining pixels (more than 72 per
cent) do not contain enough signal to reach the required s.n.r.\ even at
the largest permissible smoothing scale. The majority of the image is smoothed
at the largest possible scale of 63.8 pixels at which the dimensions of the
background kernel array (a 1$\sigma$ wide annulus surrounding the
two-dimensional Gaussian smoothing kernel computed out to $3\sigma$ radius)
equals that of the image itself. The s.n.r.\ maps (Fig.~\ref{ms1054},
right-most column), finally, give evidence of the near-constant s.n.r.\ of
all regions smoothed with kernels other than the largest one.

\subsubsection{Statistical properties of the residual image}

\begin{figure}
	\epsfxsize=0.53\textwidth
	\hspace*{-7mm} \epsffile{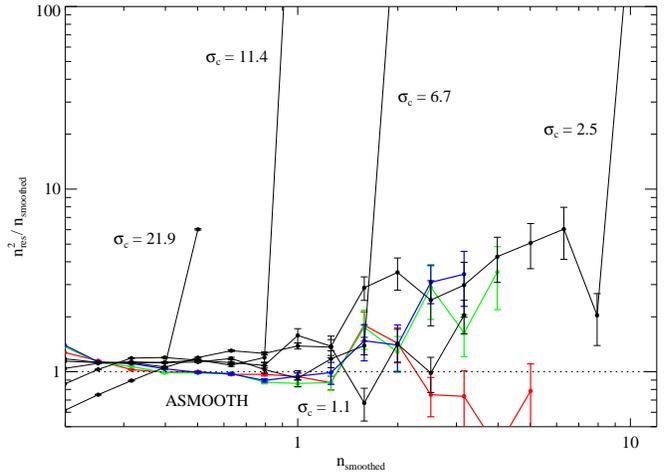}
        \caption[]{The ratio of the variance of the residual signal to the mean
          (as given by the smoothed signal) as a function of the smoothed signal
          for the example shown in Fig.~\protect\ref{ms1054}. Only bins
          containing at least 5 pixels are shown.  Also shown are Gaussian
          errors based on the number of pixels per bin. For a perfect smoothing
          algorithm the plotted ratio is unity at all values of the mean. The
          bold lines (red, green, and blue) representing the results obtained
          with {\sc asmooth} for $\tau_{\rm min}=2,\,3,$ and 4 come close to the
          ideal of a constant value of unity for most values of the mean. Only a
          few bright pixels exhibit significantly higher variance than expected
          for Poissonian statistics.  By comparison, fixed kernel smoothing (see
          the fine solid lines) results in far too high variances for all but
          the smallest kernel size.  Strong deviations from Poissonian
          statistics are observed over large portions of the image. The chosen
          fixed kernel sizes assume the values of the $1/100^{\rm
            th},\,3/100^{\rm th},\,1/10^{\rm th},\, 1^{\rm st},$ and $10^{\rm
            th}$ percentile of the distribution of {\sc asmooth} kernel sizes
          for $\tau_{\rm min}=3$ (1.1, 2.5, 6.7, 11.4, and 21.9
          pixels).}  \label{asm_csm2}
\end{figure}

The qualitative demonstration of the performance of {\sc asmooth} shown in
Fig.~\ref{ms1054} can be made more quantitative by comparing it with the results
obtained with fixed-kernel smoothing. To this end we examine the properties of
the residual images obtained by subtracting the respective smoothed image from
the observed raw image. Following Pi\~{n}a \& Puetter (1992) who introduced this
criterion in the context of Bayesian image reconstruction, we state that, for an
ideal smoothing algorithm, this residual image should contain only uncorrelated
Poissonian noise around a zero mean.  A {\em global}\/ mean of zero is
guaranteed -- within the numerical accuracy of the convolution code -- by the
requirement that any smoothing algorithm conserve counts. The requirements that
the mean also be zero {\em locally}, that the residual signal have Poissonian
variance, and that the residual signal be uncorrelated across the image are much
harder to meet. In the following we shall examine how well adaptive and fixed
kernel smoothing fulfill these requirements.

{\em Is the residual signal consistent with Poissonian noise shifted to
zero mean as expected if shot noise dominates? }

If so, the square of the noise (given by the residual signal), $n_{\rm res}^2$,
should equal the mean (given by the smoothed signal $n_{\rm smooth}$). Since the
smoothed image contains a wide range of mean values, we can test for this
condition only within bins of similar mean.  Figure~\ref{asm_csm2} shows the
ratio $n_{\rm res}^2/n_{\rm smooth}$ as a function of $n_{\rm smooth}$ for both
{\sc asmooth} and a number of fixed kernels of various sizes. Note how the
residual signal obtained with {\sc asmooth} exhibits a near-Poissonian variance
over a larger range of mean values than does the residual produced by smoothing
with a fixed kernel of essentially any size. Only the smallest fixed kernel size
tested here, $\sigma_c=1.1$ pixels, yields comparable results --- a kernel of
such small size, however, also provides essentially no smoothing.

\begin{figure}
	\epsfxsize=0.53\textwidth
	\hspace*{-7mm} \epsffile{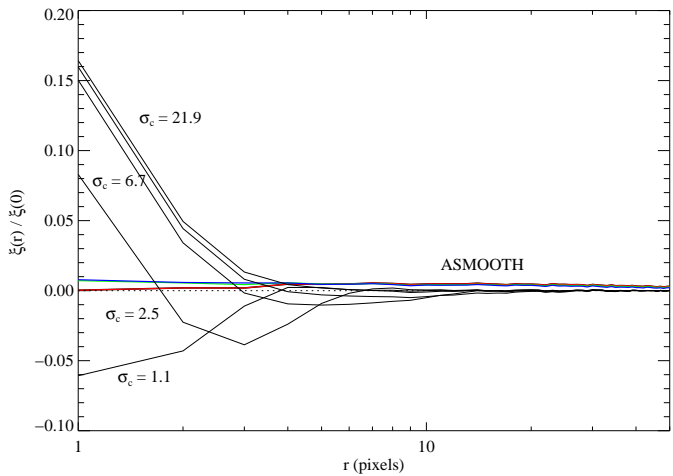}
        \caption[]{The normalized autocorrelation function
          (cf.\ Eq.~\protect\ref{acorr}) of the signal in the residual images
          obtained by subtracting the smoothed image from the original one for
          the example shown in Fig.~\protect\ref{ms1054}. For a perfect
          smoothing algorithm the autocorrelation signal is zero at all lags
          $r>0$. The almost indistuinguishable bold lines (red, green, and blue)
          representing the results obtained with {\sc asmooth} for $\tau_{\rm
            min}=2,\,3,$ and 4 come close to this ideal at the price of a slight
          positive correlation on all scales. Fixed kernel smoothing always
          results in significant spatial correlations. The chosen fixed kernel
          sizes assume the values of the $1/100^{\rm th},\,3/100^{\rm
            th},\,1/10^{\rm th},\, 1^{\rm st},$ and $10^{\rm th}$ percentile of
          the distribution of {\sc asmooth} kernel sizes for $\tau_{\rm min}=3$
          (1.1, 2.5, 6.7, 11.4, and 21.9 pixels).}  \label{asm_csm1}
\end{figure}

{\em Is the signal in the residual image spatially uncorrelated?}

Figure~\ref{asm_csm1} shows the autocorrelation function of the
residual images obtained with {\sc asmooth} and fixed kernels of
various sizes. Here we use the standard definition of the
autocorrelation $\xi$ as a function of radial lag $r$:
\begin{equation}
\xi(r) = \frac{\langle I_{\rm res}(\vec{r}\,') \,I_{\rm
              res}(\vec{r}\,'+\vec{r}\,) \rangle} {\langle I_{\rm res}
              \rangle^2} - 1
\label{acorr}
\end{equation}
where the angular brackets signify averaging over the position $\vec{r}\,'$
within the image (e.g.\ Peebles 1980). In the normalized representation of
$\xi(r)$ shown in Fig.~\ref{asm_csm1} an ideal smoothing algorithm would produce
zero signal at all lags except for a singular value of unity at $r=0$. {\sc
  Asmooth} comes close to this ideal: the signal in the residual images generated
with {\sc asmooth} is essentially uncorrelated at all scales except for a weak
($\le 2\%$) positive signal at scales smaller than about the maximum adaptive
kernel size. Note the very different result when fixed Gaussian kernels of size
$\sigma_{\mbox{\scriptsize const}}$ are used: strong spatial correlations are
observed in the residual images at all scales smaller than about the kernel
size. Only for very small kernel sizes does non-adaptive smoothing come close to
meeting the requirement that any residual signal after smoothing be spatially
uncorrelated. However, such small kernel sizes perform very poorly in the
presence of significant structure on a large range of scales
(cf.~Fig.~\ref{ms1054}, center column).

\subsection{Simulated data}
\label{simsec}

The example shown in the previous section has the advantage of using real data
but, for this very reason, does not allow the user to assess quantitatively how
the {\sc asmooth}ed image compares to the true counts distribution underlying
the noisy input image. We therefore present in this section results obtained for
a simulated data set that contains multiple extended and point-like features, as
well as a constant background component.

\begin{figure*}
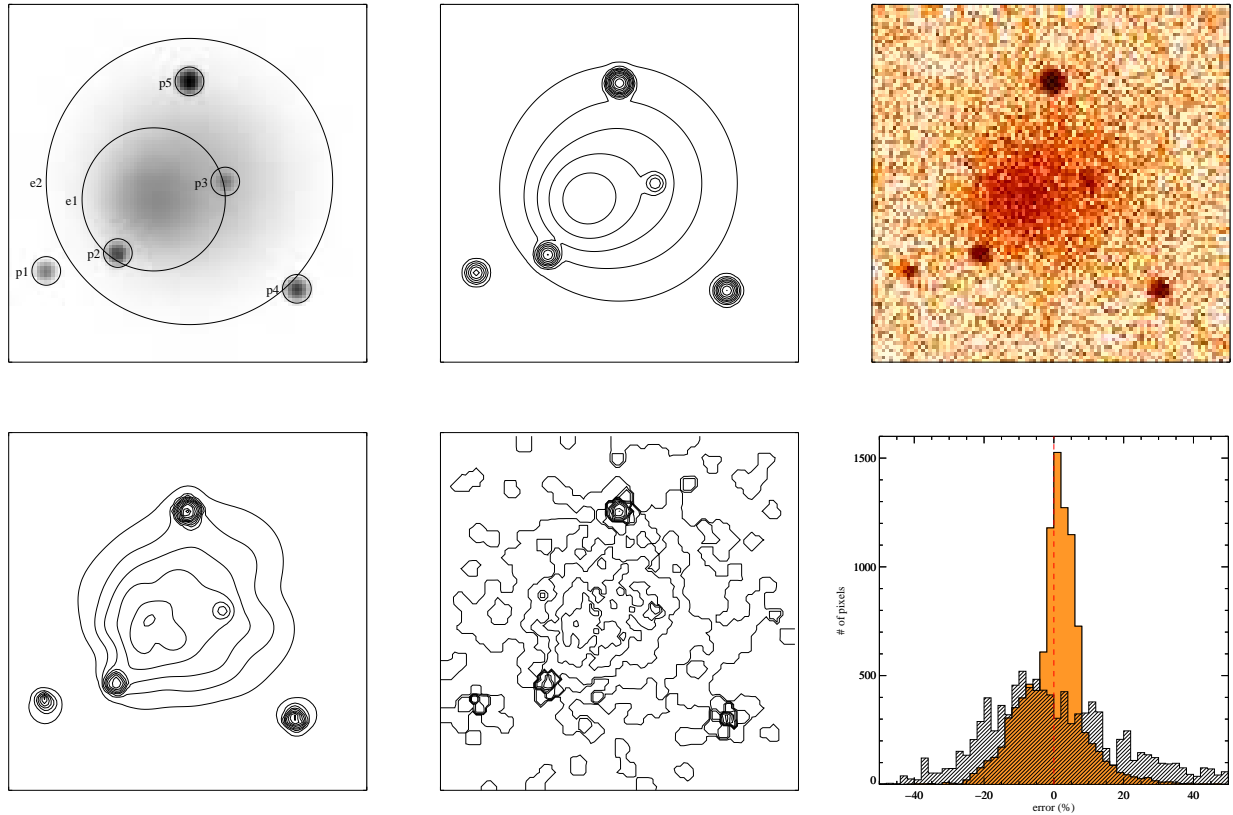

\parbox{0.33\textwidth}{
\epsfxsize=0.32\textwidth
\epsffile{asmooth_sim.fig1.epsf}}
\parbox{0.33\textwidth}{
\epsfxsize=0.32\textwidth
\epsffile{asmooth_sim.fig2.epsf}}
\parbox{0.33\textwidth}{
\epsfxsize=0.32\textwidth
\epsffile{asmooth_sim.fig3.epsf}}\mbox{}\\*[-1.5cm]
\parbox{0.33\textwidth}{
\epsfxsize=0.32\textwidth
\epsffile{asmooth_sim.fig5.epsf}}
\parbox{0.33\textwidth}{
\epsfxsize=0.32\textwidth
\epsffile{asmooth_sim.fig6.epsf}}
\parbox{0.33\textwidth}{
\epsfxsize=0.32\textwidth
\epsffile{asmooth_sim.fig7.epsf}}\mbox{}\\*[-1.5cm]
\caption{\label{simplot}{\bf Top row.} Left: The model used in our simulation, a
  combination of two extended, four compact features, and a spatially invariant
  background component, designed to mimic an astrophysical X-ray image of point
  sources superimposed on asymmetric diffuse emission from, e.g.\ a cluster of
  galaxies. Center: Iso-intensity contours for the same model. Right: input
  image obtained by applying Poisson noise to the model. {\bf Bottom row.} Left:
  Iso-intensity contours of the {\sc asmooth}ed image obtained with a Gaussian
  kernel and $\tau_{\rm min}=3$ -- the contour levels are the same as above for
  the model image. Center: Iso-intensity contours of an adaptively {\em
    binned}\/ image as obtained with the WVT algorithm (Diehl \& Statler 2005)
  for an s.n.r.\ target value of 5 -- the contour levels are again the same as
  for the model image. Note that, unlike {\sc asmooth}, WVT does not correct for
  a local background in its s.n.r.\ estimation. Right: Histogram of relative
  errors, defined as (result -- model)/model, in percent for {\sc Asmooth}
  (solid orange fill) and the WVT adaptive binning code (hatched).  }
\end{figure*}

Figure~\ref{simplot} summarizes the characteristics of the model used for our
simulation (top row). The same figure shows, in the bottom row, results obtained
with {\sc Asmooth} for $\tau_{\rm min}=3,\,\tau_{\rm max}=\tau_{\rm min}+1$, and
with the adaptive {\em binning}\/ algorithm WVT (Diehl \& Statler 2005) for a
target s.n.r.\ value of 5. Note how the requirement of a minimal s.n.r.\ {\em
  above the local background}\/ (Eqs.~4 and 5) causes {\sc Asmooth} to reduce
noise much more aggressively than WVT in spite of a nominally lower
s.n.r.\ target value.  As demonstrated by the error distribution depicted in the
final panel of Fig.~\ref{simplot} the {\sc asmooth} residual image exhibits a
near Gaussian error distribution around zero mean, in addition to featuring
Poissonian variance and being spatially uncorrelated
(Section~\ref{ms1054}). {\sc Asmooth}ed images can thus be considered to be fair
representations of the `true' input image.

\section{Summary}

We describe {\sc asmooth}, an efficient algorithm for adaptive kernel smoothing
of two-dimensional image data. {\sc Asmooth} determines the local size of the
kernel from the requirement that, at each position within the image, the
signal-to-noise ratio of the counts under the kernel, and above the background,
must reach (but not exceed greatly) a certain preset minimum. Qualitatively,
this could be called the `uniform significance approach'. As a consequence of
this criterion, noise is heavily suppressed and real structure is enhanced
without loss of spatial resolution.  Due to the choice of boundary conditions,
the algorithm preserves the total number of counts in the raw data.  The {\sc
  asmooth}ed image is a fair representation of the input data in the sense that
the residual image is consistent with pure noise, i.e., it the residual
possesses Poissonian variance and is spatially uncorrelated. As demonstrated by
the results obtained for simulated input, the adaptively smoothed images created
by {\sc Asmooth} are fair representations of the true counts distribution
underlying the noisy input data.

Since the background is accounted for in the assessment of any structure's
s.n.r., a feature that distinguishes {\sc asmooth} from most other adaptive
smoothing or adaptive binning algorithms, the interpretation of the adaptively
smoothed image is straightforward: all features in the {\sc asmooth}ed image are
equally significant at the scale of the kernel used in the smoothing. A map of
these smoothing scales is returned by {\sc asmooth} together with the smoothed
image. Note that these are the {\em smallest}\/ scales at which real features
reach the selected s.n.r.\ threshold. Consequently, background regions never
meet the s.n.r.\ criterion and are smoothed at the largest possible scale for
which the kernel size equals the size of the image. Such regions of insufficient
signal are easily flagged using an s.n.r.\ map which is also provided by our
algorithm.

We emphasize that our definition of ``signal'' as meaning ``signal {\em above
  the (local) background}'' was chosen because many, if not most, real-life
appplications do not provide the user with a priori knowledge of the intensity
of the background or any spatial background variations. Also, superpositions of
features (such as the point sources on top of diffuse emission in our Chandra
ACIS-S example) require a {\em local}\/ background estimate if the intrinsic
signal of features on very different scales is to be assessed reliably. The
approach taken by {\sc Asmooth} is thus intrinsically very different from the
one implemented in, for instance, the adaptive binning algorithm WVT (Diehl \&
Statler 2005) and the difference in the resulting output is accordingly dramatic
(see Section~\ref{simsec}).

{\sc Asmooth} is being used extensively in the analysis of astronomical X-ray
imaging data gathered in a wide range of missions from ROSAT to XMM-Newton, and
has become the analysis tool of choice for the display of high-resolution images
obtained with the Chandra X-ray observatory. Recent examples illustrating {\sc
  asmooth}'s performance can be found in Ebeling, Mendes de Oliveira \& White
(1995), Brandt, Halpern \& Iwasawa (1996), Hamana et al.\ (1997), Ebeling et
al.\ (2000), Fabbiano, Zezas \& Murray (2001), Krishnamurthi et al.\ (2001),
Karovska et al.\ (2002), Bauer et al.\ (2002), Gil-Merino \& Schindler (2003),
Heike et al.\ (2003), Rasmussen, Stevens \& Ponman (2004), Clarke, Blanton \&
Sarazin (2004), Ebeling, Barrett \& Donovan (2004), or Pratt \& Arnaud (2005), 
as well as in many Chandra press releases (\mbox{\tt{http://chandra.harvard.edu/press/}}).

Under development is an improved version of the algorithm which accounts for
Poisson statistics using the analytic approximations of Ebeling (2003) to allow
proper treatment of significant negative features, such as absorbed regions,
detector chip gaps, or instrument elements (e.g., the window support structure
of the ROSAT PSPC) obscuring part of the image.

{\sc Asmooth} is written in the IDL programming language. The source code is
available on request from ebeling{@}ifa.hawaii.edu.  A $C++$ version of an early
version of the code, called {\sc csmooth}, is part of CIAO, the official suite
of data analysis tools for the Chandra X-ray observatory, and is not identical to
the algorithm described here.

\section*{Acknowledgements}

We thank all {\sc Asmooth} users in the community who took the time to provide
comments and criticism that have resulted in a dramatic improvement of the
algorithm and its implementation since its original inception in
1995. Discussions with Barry LaBonte about the statistical properties of
residual shot noise. Thanks also to our referee, Herv\'e Bourdin, for
constructive criticism and many useful suggestions that helped to improve this
paper significantly. HE gratefully acknowledges partial financial support from
many sources over the years during which {\sc Asmooth} was developed and
improved, among them a European Union EARA Fellowship, SAO contract SV4-64008,
as well as NASA grants NAG 5-8253 and NAG 5-9238. DAW acknowledges support from
PPARC.


\begin{thebibliography}{9}
\bibitem{bauer} Bauer F.E.\ et al. 2002, AJ, 123, 1163
\bibitem{brandt} Brandt W.N., Halpern J.P.\ \& Iwasawa K. 1996, MNRAS, 281, 687
\bibitem{hcg} Ebeling H., Mendes de Oliveira C.\ \& White D.A. 1995,
	MNRAS, 277, 1006
\bibitem{cap} Cappellari M.\ \& Copin Y. 2003, MNRAS, 342, 345
\bibitem{clarke} Clarke T.E., Blanton E.L.\ \& Sarazin C.L. 2004, 616, 178
\bibitem{diehl} Diehl S.\ \& Statler T.S. 2005, MNRAS, submitted
\bibitem{warps} Ebeling H.\ et al. 2000, ApJ, 534, 133
\bibitem{he} Ebeling H.\ 2003, MNRAS, 340, 1269
\bibitem{0717} Ebeling H., Barrett E.\ \& Donovan, D. 2004, ApJ, 609, L49
\bibitem{peppi} Fabbiano G., Zezas A.\ \& Murray S.S. 2001, ApJ, 554, 1035
\bibitem{gil} Gil-Merino R.\ \& Schindler S. 2003, A\&A, 408, 51
\bibitem{hamana} Hamana T., Hattori M., Ebeling H., Henry J.P., Futamase T., 
	Shioya Y. 1997, ApJ, 484, 574
\bibitem{hei} Heike K., Awaki H., Misao Y., Hayashida K., Weaver K.A. 2003
	ApJ, 591, 99L
\bibitem{huang} Huang Z.\ \& Sarazin C. 1996, ApJ, 461, 622
\bibitem{jeltema} Jeltema T.E., Canizares C.R., Bautz M.W., Malm M.R.,
   	Donahue M., Garmire G.P. 2001, ApJ, 562, 124
\bibitem{kar} Karovska M., Fabbiano G., Nicastro F., Elvis M., Kraft R.P.,
 	Murray S.S. 2002, ApJ, 2002, 577, 114
\bibitem{kri} Krishnamurthi A., Reynolds C.S., Linsky J.L., Martin E., 
	Gagn\'{e} M. 2001, AJ, 121, 337
\bibitem{paff} Lorenz H., Richter G.M., Capaccioli M., Longo G. 1993, A\&A, 
               277, 321
\bibitem{merritt}  Merritt D.\ \& Tremblay B. 1994, AJ, 108, 514
\bibitem{peebles} Peebles P.J.E. 1980, {\em The Large-Scale Structure of the Universe},
                  Princeton, Princeton University Press
\bibitem{pina} Pi\~{n}a R.K.\ \& Puetter R.C. 1992, PASP, 104, 1096
\bibitem{pisani_1} Pisani A. 1993, MNRAS, 265, 706
\bibitem{pisani_2} Pisani A. 1996, MNRAS, 278, 697
\bibitem{pratt} Pratt G.W.\ \& Arnaud M. 2005, A\&A, 429, 791
\bibitem{ras} Rasmussen J., Stevens I.R.\ \& Ponman T.J. 2004, MNRAS, 354, 259
\bibitem{richter} Richter G.M., Bohm P., Lorenz H., Priebe A.  1991,
                  Astron.~Nachr., 312, 346
\bibitem{sanders} Sanders J.S.\ \& Fabian A.C. 2001, MNRAS, 325, 178
\bibitem{silverman} Silverman B.W. 1986, Density Estimation for Statistics and 
                    Data Analysis, Chapman \& Hall, London
\bibitem{starck} Starck J.-L.\ \& Pierre M. 1998, A\&AS, 128, 397
\bibitem{thomplso} Thompson A.M. 1990, A\&A, 240, 209
\bibitem{vio}      Vio R., Fasano, G., Lazzarin M., Lessi, O. 1994, A\&A, 289,
		   640
\end{thebibliography}
\end{document}